\documentclass[sn-mathphys]{sn-jnl}

\usepackage{graphicx,subfigure}
\usepackage{color}
\usepackage{soul}
\newcommand{\Rho}{\mathrm{P}}

\jyear{2022}%

\theoremstyle{thmstyleone}%
\theoremstyle{thmstyletwo}%

\theoremstyle{thmstylethree}%

\begin{document}

\title[Expected Performance of iXRD in the Orbital Background Radiation]{A Simulation Study for the Expected Performance of Sharjah-Sat-1 payload improved X-Ray Detector (iXRD) in the Orbital Background Radiation}


\author*[1]{\fnm{Ali} \sur{M. Alt{\i}ng\"un}}\email{aaltingun@sabanciuniv.edu}
\author*[1]{\fnm{Emrah} \sur{Kalemci}}\email{ekalemci@sabanciuniv.edu}

\author[1]{\fnm{Efe} \sur{\"Oztaban}}\email{efeoztaban@sabanciuniv.edu}

\affil*[1]{\orgdiv{Faculty of Engineering and Natural Sciences}, \orgname{Sabanci\i\ University}, \orgaddress{\street{Orta Mah. Tuzla}, \city{Istanbul}, \postcode{34956}, \country{Turkey}}}




\abstract{Sharjah-Sat-1 is a 3U cubesat with a CdZnTe based hard X-ray detector, called iXRD (improved X-ray Detector) as a scientific payload with the primary objective of monitoring bright X-ray sources in the galaxy. We investigated the effects of the in-orbit background radiation on the iXRD based on Geant4 simulations. Several background components were included in the simulations such as the cosmic diffuse gamma-rays, galactic cosmic rays (protons and alpha particles), trapped protons and electrons, and albedo radiation arising from the upper layer of the atmosphere. The most dominant component is the albedo photon radiation which contributes at low and high energies alike in the instrument energy range of 20 keV - 200 keV. On the other hand, the cosmic diffuse gamma-ray contribution is the strongest between 20 keV and 60 keV in which most of the astrophysics source flux is expected. The third effective component is the galactic cosmic protons. The radiation due to the trapped particles, the albedo neutrons, and the cosmic alpha particles are negligible when the polar regions and the South Atlantic Anomaly region are excluded in the analysis. The total background count rates are $\sim$0.36 and $\sim$0.85 counts/s for the energy bands of 20 - 60 keV and 20 - 200 keV, respectively.	We performed charge transportation simulations to determine the spectral response of the iXRD and used it in sensitivity calculations as well. The simulation framework was validated with experimental studies. The estimated sensitivity of 180 mCrab between the energy band of 20 keV - 100 keV indicates that the iXRD could achieve its scientific goals.}




\maketitle

\section{Introduction}\label{intro}

\begin{figure}[ht]
	\includegraphics[width=0.9\textwidth]{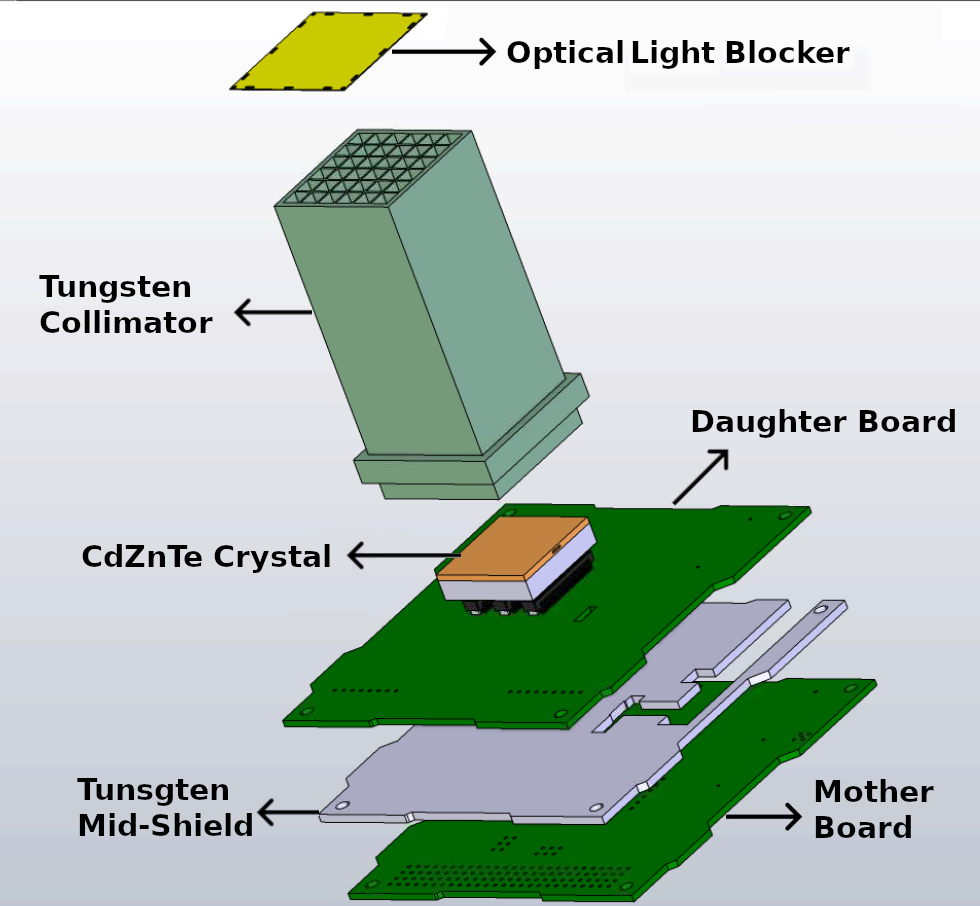}
	\caption{A plain CAD drawing of the iXRD system. Daughterboard carries the RENA and associated power supplies and motherboard carries the digital control electronics.}
	\label{fig:ixrdCAD}       
\end{figure}

Sharjah-Sat-1 is a 3U cubesat X-ray satellite being developed in collaboration with the  Sharjah  Academy for  Astronomy,  Space  Sciences, and Technology, the University of Sharjah, Istanbul Technical University, and  Sabanci  University. Its scientific payload, the iXRD (improved X-Ray Detector), is a CdZnTe based hard X-ray detector with the main objective of monitoring bright X-ray sources \cite{Kalemci2021}. The iXRD is an improved version of XRD (X-Ray Detector) that has been employed as a scientific payload for the 2U BeEagleSAT cubesat\cite{Kalemci2013}. While the XRD served as a demonstrator, the iXRD is a system with point source observing capability and better spectral and noise performance thanks to the improvements made in the readout electronics and mechanical design and the addition of a collimator \cite{Kalemci2022}. The iXRD has 2.54$\times$2.54$\times$0.5 cm$^{3}$ pixellated CdZnTe crystal produced by eV Products (Kromek). It consists of 256 pixels (16 $\times$ 16) with 1.6 mm pitch size and a planar cathode. 
The energy range is from 20 keV to 200 keV, limited by the readout electronics. For readout, RENA 3b ASIC (application-specific integrated circuit) with 36 channels is used\cite{Tumer08}. Therefore, some of the pixels are connected into groups forming single (one pixel), small (6 pixels),  medium (8-9 pixels), and large (10-12 pixels) channels. There is a square hole Tungsten collimator over the top of the crystal. Its structure also encircles the crystal for additional background protection. The collimator provides a 4.26$^{\circ}$ field of view (FoV) which enables the iXRD to observe point sources and reduces the cosmic X-ray background substantially. In addition, an aluminum plate of 0.3 mm is positioned on top of the collimator as an optical light blocker. Finally, a 2 mm thick tungsten plate (called back-shield) is placed under the crystal serving as a passive shielding against all radiation, especially the components arising at Earth's albedo. A simple CAD drawing of the iXRD components is illustrated in Fig.~\ref{fig:ixrdCAD}. The details of the design and on-ground performance of the detector are given in \cite{Kalemci2022}.

The cubesat is planned to be launched into a near-polar sun-synchronous orbit (SSO) with an altitude of 500 - 600 km. The harsh radiation environment in the low Earth orbit (LEO) has crucial effects on the operation of spacecrafts. High-energy particles and photons decrease the observation performance. Moreover, the charged particles cause radiation damage to the electronic systems and may degrade CdZnTe crystal properties in time. Therefore, understanding the in-orbit radiation effects plays an important role in the performance of the iXRD. The background radiation consists of two main components, prompt and delayed radiation. The prompt radiation is due to the trapped charged particles by the Earth's magnetic field, cosmic diffuse gamma-rays, galactic cosmic rays, albedo radiation originating from the interactions of cosmic particles with the atmosphere. Long-term irradiation of the cosmic and trapped protons can produce radioactive isotopes within the material of the satellites. The emission from the induced isotopes is called delayed background radiation. Considering the size of Sharjah-Sat-1 and the mission lifetime of 1-2 years, background due to the activation has not been considered in this work. \\

We performed several Monte Carlo simulations to estimate the background effects by using GEANT4 software package \cite{Agostinelli2003}, which allows us to calculate energy depositions in the CdZnTe crystal by the incoming radiation. In this way, we determined the contributions of each component to the total background level. Also, an optimization study for the back-shield design to mitigate the effects of the albedo radiation has been carried out. Finally, by using the simulated background rates and charge transportation simulation results, we calculated the sensitivity of the iXRD.

\section{Components of the background radiation environment in the LEO}
\label{sec:background_components}
In our work, we considered several background components such as the cosmic diffuse gamma-rays, galactic cosmic rays, trapped particles due to the Earth's magnetic field, and albedo particles and photons originating from the interactions of cosmic particles with the atmosphere. In the following sections, we describe the spectrum of each component that is used as an input for the Geant4 simulations. 

\subsection{Cosmic diffuse gamma-rays}
\label{sec:CDGR_ch}
The first evidence for the cosmic diffuse gamma-rays (CDGR) was reported by a rocket-borne large area Geiger counters in 1962\cite{Giacconi1962}. At soft X-ray energies, primary contributors are active galactic nuclei (AGNs) \cite{Luo2016} and X-ray Binaries (XRB) in the distant galaxies \cite{Lehmer2016}, while at hard X-ray and gamma-ray energies, the contributions are mainly due to the AGNs components, supernovas and blazars\cite{Gruber1999TheSO}. The differential flux of the CDGR in the LEO is given by Eq. \ref{eq:CDGR_eq} for the energies between 10 keV to 100 MeV\cite{Gruber1999TheSO}. The given flux is in the unit of particles$/$cm$^{2}$$/$s$/$sr$/$keV.

\begin{equation}
	\frac{f\big(E\big)}{dE} =
	\begin{cases}
		\text{$\quad$  7.877$\times$ E$^{-1.29}$ exp$\big( \frac{-E}{41.13}\big)$ $\qquad$ for 3 keV $\leq$ E $\leq$ 60 keV}\\
		\\
		\text{$\quad$  4.32$\times$10$^{-4}$ $\big($$\frac{E}{60}\big)^{-6.5} $}\\
		\text{$\quad$ + 8.4$\times$10$^{-3}$ $\big($$\frac{E}{60}\big)^{-2.58}$ $\qquad$  $\quad$ for E$\geq$ 60 keV}\\
		\text{$\quad$ + 4.8$\times$10$^{-4}$ $\big($$\frac{E}{60}\big)^{-2.05}$}
	\end{cases}       
	\label{eq:CDGR_eq}
\end{equation}


\subsection{Trapped particles}
\label{sec:trped_ch}
Unlike the other terrestrial planets in the Solar System, there is an immense magnetic field around the Earth which serves as a strong shield for the charged particles originating from the Sun and outer space. The particles streaming toward the Earth are trapped by the magnetic field and move along the field lines for long periods. Although the first significant studies on the existence of the trapped charged particles were carried out towards the end of the 19th century, it was first confirmed by the \emph{Explorer 1} spacecraft equipped with a Geiger Muller tube in 1958 \cite{VanAllen1958}. The trapped charged particle zones are called Van Allen radiation belts. The Van Allen belts consist of the inner belt, which has a population of MeV protons and tens to hundreds of keV electrons, and the outer belt which is mostly dominated by MeV electrons \cite{Li2019}. The population of the trapped particles is not stable and varies in the orbit due to effects such as solar activity and the local changes in the Earth's magnetic field. The South Atlantic Anomaly (SAA) is the area where the magnetic field lines dip down to low altitudes, which is surmised to be caused by the complicated motion of the molten metals in the Earth's outer core and the tilt of the Earth's magnetic field axis with respect to the rotation axis. The trapped proton population is quite high inside the SAA region. Rigidity of a charged particle is a measure of the particle stiffness against the magnetic field to be bent. Particles with the same rigidity, charge, and initial conditions will follow the same path under a static magnetic field. For each coordinate on the Earth, there is a rigidity value, called geomagnetic cut-off rigidity, that describes the access of a charged particle to the Earth. Particles with rigidity less than the cut-off rigidity can not penetrate the magnetosphere and are deflected. Due to the low cut-off rigidities in the polar regions, the trapped electron fluxes can increase substantially at high latitudes. Since regular science observations are not possible due to high particle fluxes, the iXRD will be inoperative while passing through the SAA and the polar regions. 

\subsubsection{Trapped electrons}
\begin{figure}[ht]
	\includegraphics[width=0.9\textwidth]{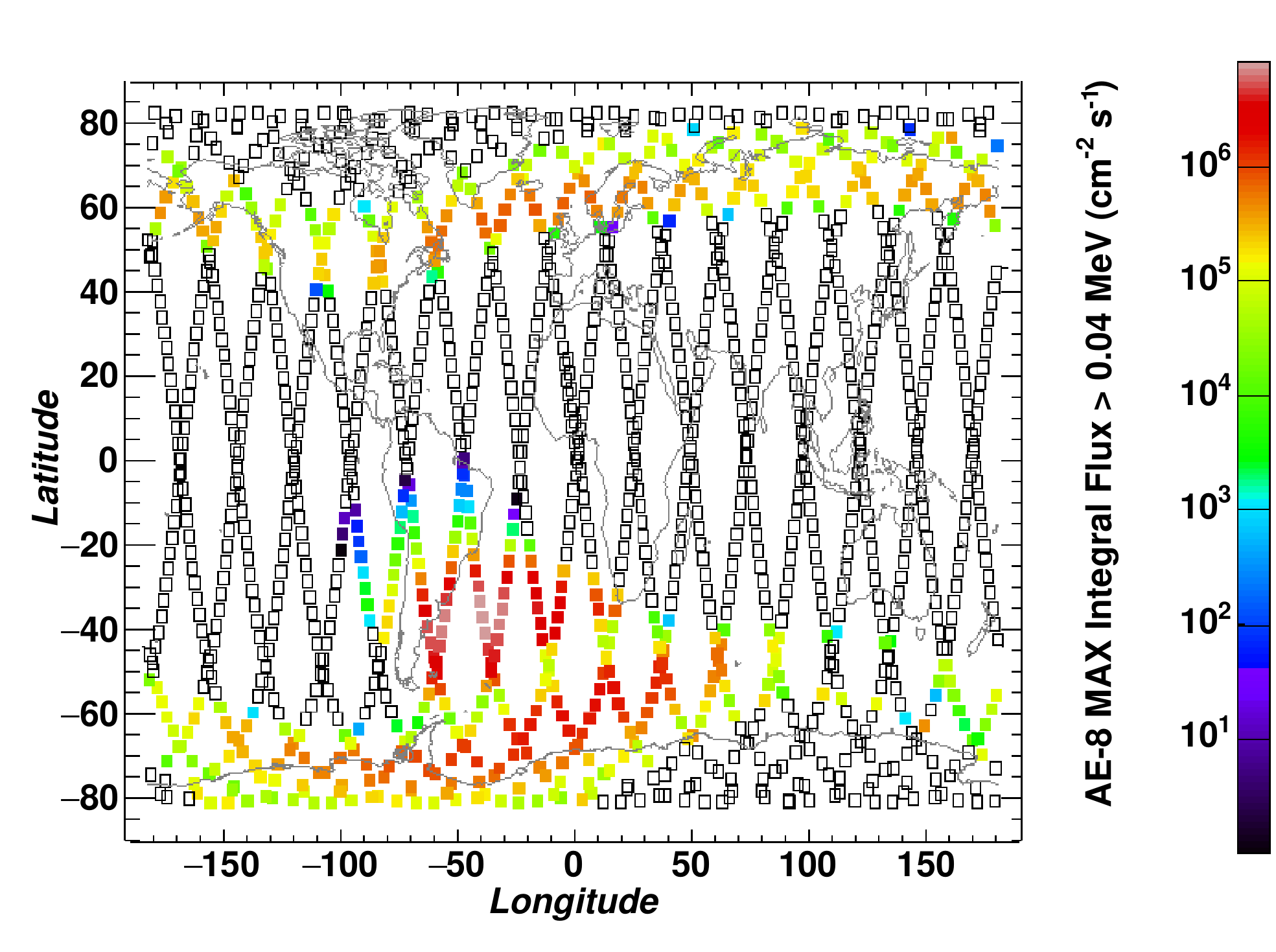}
	\caption{The average trapped electron flux to which Sharjah-Sat-1 is exposed during one-day orbital trajectory.}
	\label{fig:Electron_SAA_Polar_Regions}       
\end{figure}

\begin{figure}[ht]
	\includegraphics[width=0.9\textwidth]{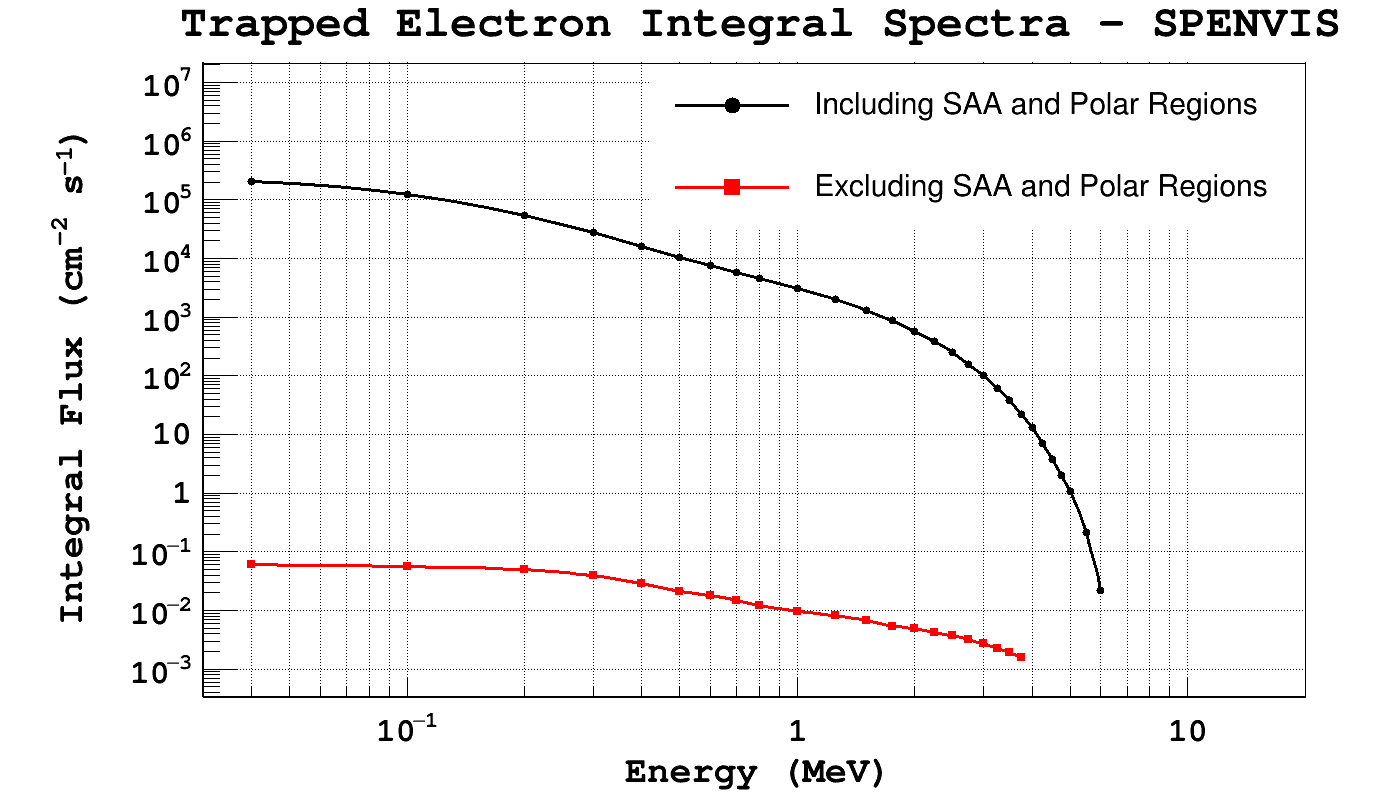}
	\caption{The integral spectra for the orbit-averaged trapped electrons with AE8 MAX model including/excluding count rates inside the SAA and polar regions. }
	\label{fig:Electron_Spectra_SAA_Polar_Regions}       
\end{figure}

Trapped electron flux averaged over one year mission was obtained from SPENVIS by using the AE8 MAX model in the energy range of 40 keV to 7 MeV. Average trapped electron exposure on the cubesat along the 1-day orbital trajectory is shown in Fig.~\ref{fig:Electron_SAA_Polar_Regions}. The squares in the figure indicate the orbital positions of the cubesat and each position corresponds to approximately 60 seconds of flight time. The average electron flux is extremely high in the SAA region and at high altitudes. The orbit-averaged energy spectra for the trapped electrons excluding and including the high count rate areas are given in Fig.~\ref{fig:Electron_Spectra_SAA_Polar_Regions}. It can be easily deduced from the spectra that the effect of trapped electrons is considerably reduced when the regions with high background rates are excluded in the planning of observations. 

\subsubsection{Trapped protons}
\begin{figure}[ht]
	\includegraphics[width=0.9\textwidth]{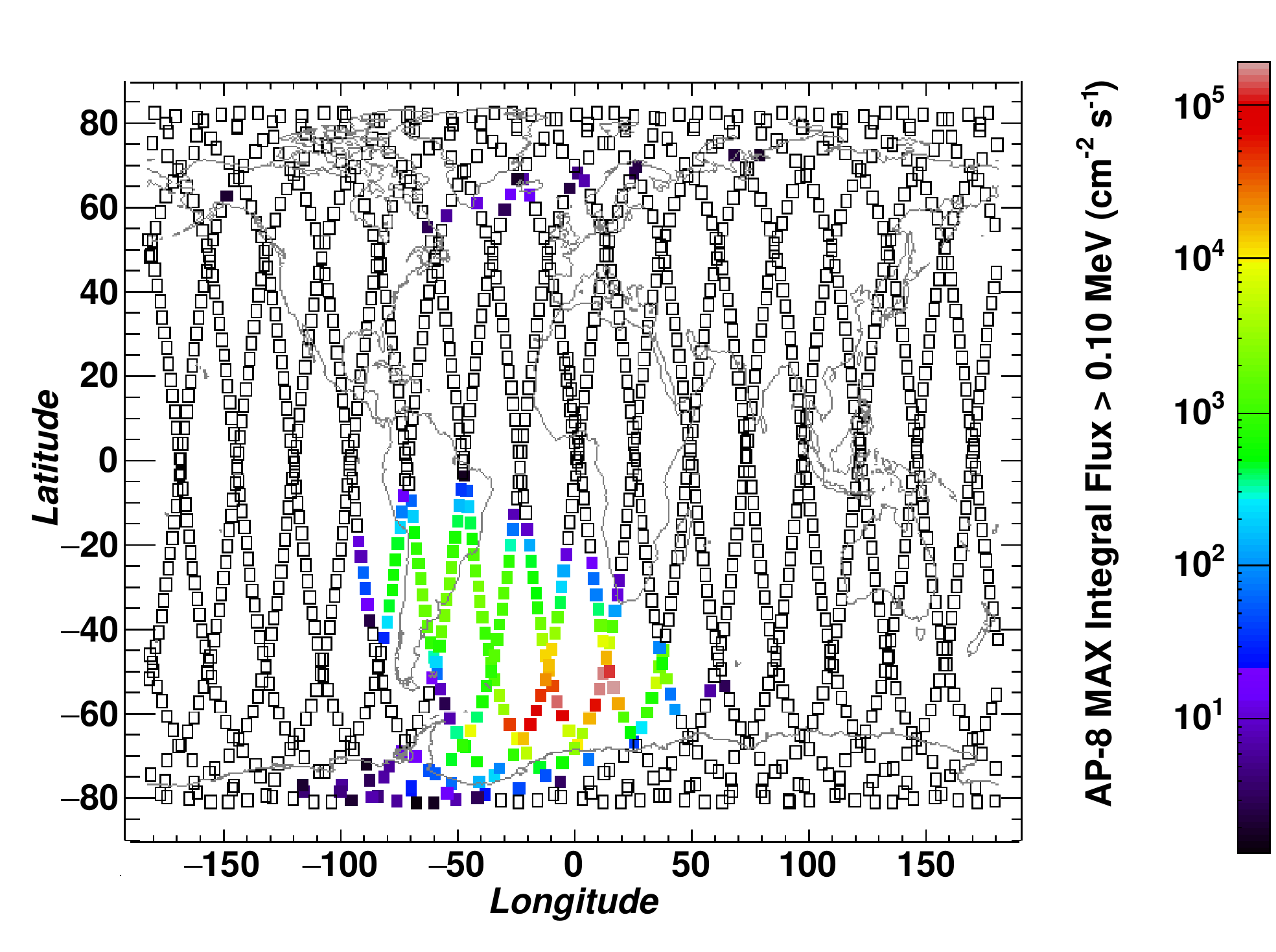}
	\caption{The averaged trapped proton flux to which Sharjah-Sat-1 is exposed during one-day orbital trajectory.}
	\label{fig:Proton_SAA_Polar_Regions}       
\end{figure}

\begin{figure}[ht]
	\includegraphics[width=0.9\textwidth]{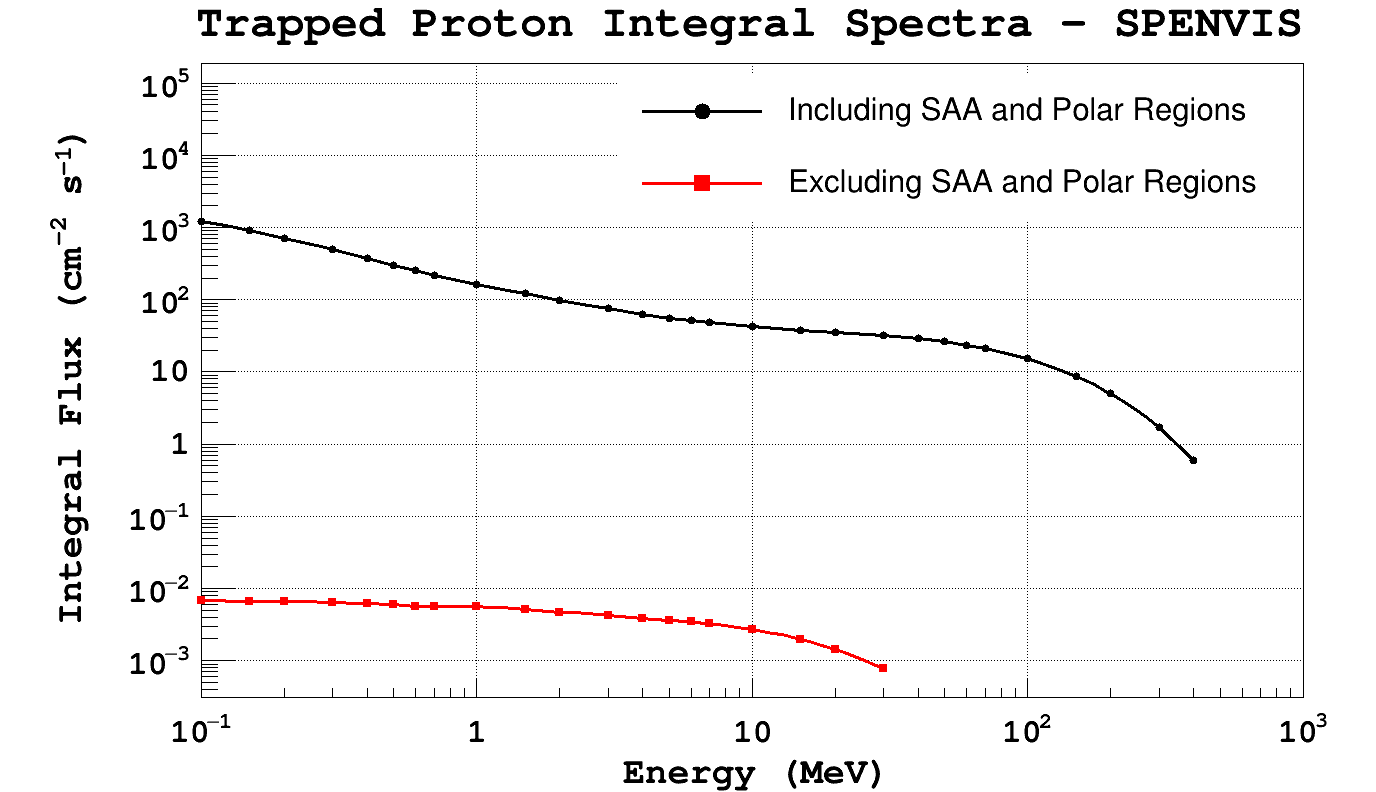}
	\caption{The integral spectra for the orbit-averaged trapped protons with AP8 MIN model including/excluding count rates inside the SAA and polar regions. }
	\label{fig:Protons_Spectra_SAA_Polar_Regions}       
\end{figure}

Average proton flux over one year mission was calculated by using the model for the trapped proton population at minimum solar activity, AP8 MIN \cite{AP8}. The energy range of the trapped protons is from 100 keV to 400 MeV in the LEO. In Fig.~\ref{fig:Proton_SAA_Polar_Regions}, average trapped proton exposure on the cubesat along the 1-day orbital trajectory is illustrated and one can see that the trapped protons are highly populated inside the SAA region. Fig.~ \ref{fig:Protons_Spectra_SAA_Polar_Regions}shows the orbit-averaged trapped proton flux and the background flux significantly drops when the SAA area is excluded. 


\subsection{Galactic cosmic rays}
\label{sec:GCR_ch}
\begin{figure}[ht]
	\includegraphics[width=0.9\textwidth]{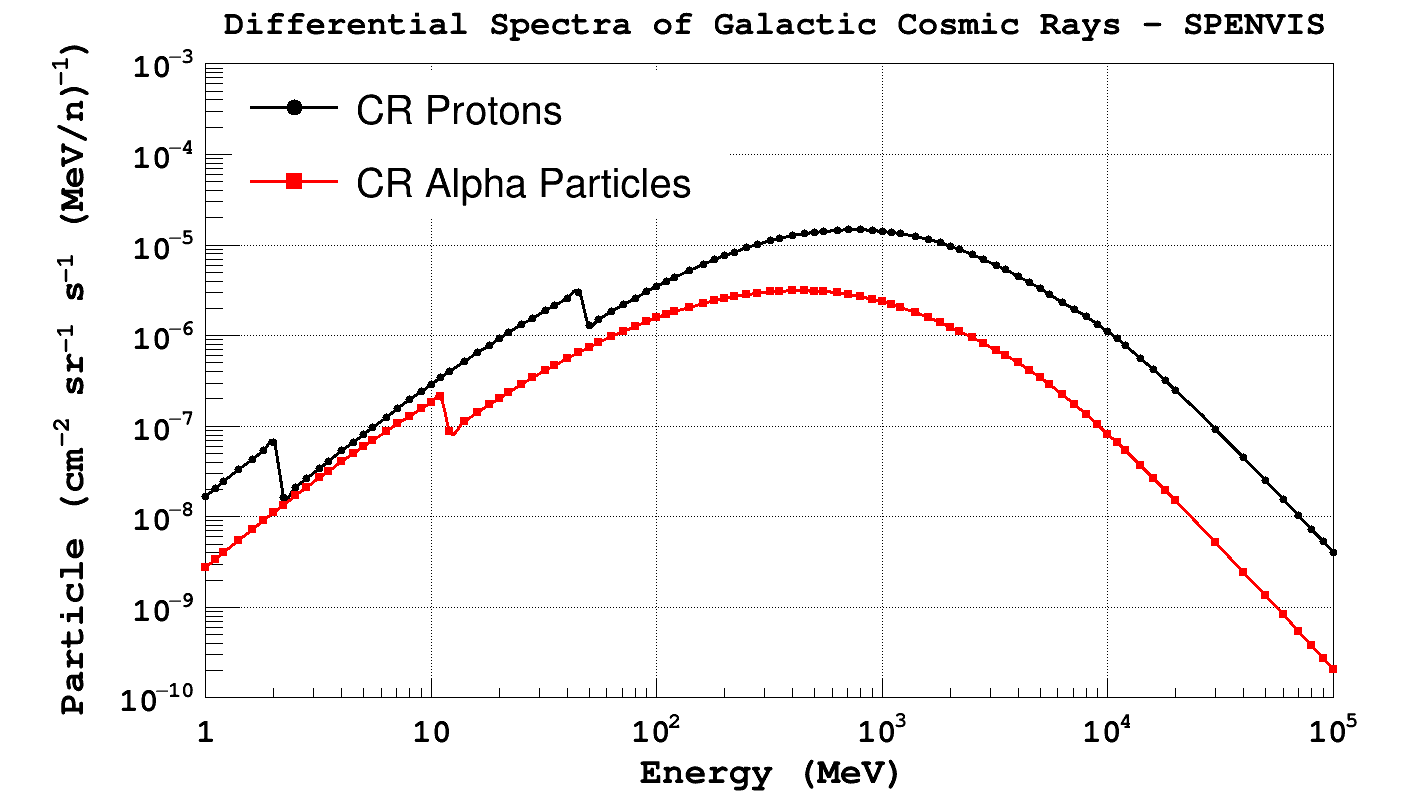}
	\caption{The differential spectra for the orbit-averaged cosmic hydrogen ions (protons) and helium ions (alpha particles) calculated by using ISO-15390 model in SPENVIS. The discontinuities in the spectra could be related with the lag of the flux variations of the GCRs relative to the solar activity variations.}
	\label{fig:GCR_particle_spectra}       
\end{figure}

Galactic cosmic rays (GCRs) originate from the outside of the Solar System. GCRs consist of approximately 98$\%$ high energetic, completely ionized nuclei ranging from hydrogen to uranium and around 2$\%$ electrons and positrons. The nuclei population consists of $\sim$87$\%$ hydrogen ions (protons), $\sim$12$\%$ helium nuclei (alpha particles), and $\sim$1$\%$ of much heavier nuclei \cite{Simpson1983}. In this work, we only considered the hydrogen ions and helium nuclei. The fluxes for the GCRs were retrieved from SPENVIS and the ISO-15390 standard model was implemented. The GCRs are considerably affected by solar winds. We used the following parameters for the ISO-15390 model: mission epoch for solar activity, magnetic shielding on, all directions, stormy magnetosphere, St$\o$rmer with eccentric dipole method, and magnetic field moment unchanged. Fig.~\ref{fig:GCR_particle_spectra} shows the differential fluxes for cosmic protons and alpha particles averaged over the spacecraft orbit obtained by the ISO-15390 model\footnote{\url{https://www.spenvis.oma.be/help/background/gcr/gcr.html\#ISO}} for the SSO orbit at 550 km altitude.

\subsection{Albedo Radiation}
\label{sec:albedo_particles}
The atmosphere of the Earth obstructs energetic particles from the Sun and outer space. From the outer layers, some fraction of the incident particles is reflected as well as create secondary (albedo) particles contributing to the radiation environment in the Earth's orbits. In this paper, we considered two secondary components, albedo photons, and neutrons. The given fluxes are in the unit of particles$/$cm$^{2}$$/$s$/$sr$/$keV.

\subsubsection{Albedo photons}
\label{sec:albedo_photon_ch}
Albedo photons are produced as a result of cosmic ray interactions with the Earth's atmosphere, as well as the reflection of the CDGR from the atmosphere. The photons with energies above 50 MeV are produced by the decay of mesons due to hadronic interactions, while the photons with lower energies ($<$ 50 MeV) are the production of bremsstrahlung of cosmic electrons and positrons with the atmospheric atoms. The details of the model implemented in this work are given in \cite{Sarkar2011}. The energy range for the albedo gamma-rays is considered between 10 keV and 100 MeV, as with the CDGR.

\begin{equation*}
	\frac{f\big(E\big)}{dE} =
	\begin{cases}
		\text{$\quad$  $\frac{1.87\times10^{-2}}{\big(\frac{E}{33.7}\big)^{-5.0} + \big(\frac{E}{33.7}\big)^{1.72}}$ $\qquad$ for E $\leq$ 200 keV}\\
		\\
		\text{$\quad$  1.01$\times$10$^{-4}$ $\big($$\frac{E}{MeV}\big)^{-1.34} $$\qquad$  $\quad$ for 200 keV $\leq$ E $\leq$ 20 MeV}\\
		\\
		\text{$\quad$  7.29$\times$10$^{-4}$ $\big($$\frac{E}{MeV}\big)^{-2.0}$ $\qquad$  $\quad$ for E $\geq$ 20 keV}\\
	\end{cases}       
	\label{eq:Albedo_photon_eq}
\end{equation*}


\subsubsection{Albedo neutrons}
\label{sec:albedo_neutron_ch}
Another component originating from the atmosphere is the albedo neutrons. Bombardment of the cosmic rays on the atmospheric nuclei produces neutron emission which contributes the background radiation environment in the LEO. The energy range for the albedo neutron is 10 keV - 1 GeV in the simulations. The spectrum for the albedo neutrons is presented as \cite{Sarkar2011}:

\begin{equation*}
	\frac{f\big(E\big)}{dE}=
	\begin{cases}
		\text{$\quad$  9.98$\times$10$^{-8}$ $\big($$\frac{E}{GeV}\big)^{-0.5} $$\qquad$  $\quad$ for 10 keV $\leq$ E $\leq$ 1 MeV}\\
		\\
		\text{$\quad$  3.16$\times$10$^{-9}$ $\big($$\frac{E}{GeV}\big)^{-1.0} $$\qquad$  $\quad$ for 1 MeV $\leq$ E $\leq$ 100 MeV}\\
		\\
		\text{$\quad$  3.16$\times$10$^{-10}$ $\big($$\frac{E}{GeV}\big)^{-2.0}$ $\qquad$  $\quad$ for 100 MeV $\leq$ E $\leq$ 100 GeV}\\
	\end{cases}       
	\label{eq:Albedo_neutron_eq}
\end{equation*}


\section{Simulations}
\label{sec:Simulations}
\begin{figure}[ht]
	\includegraphics[width=0.8\textwidth]{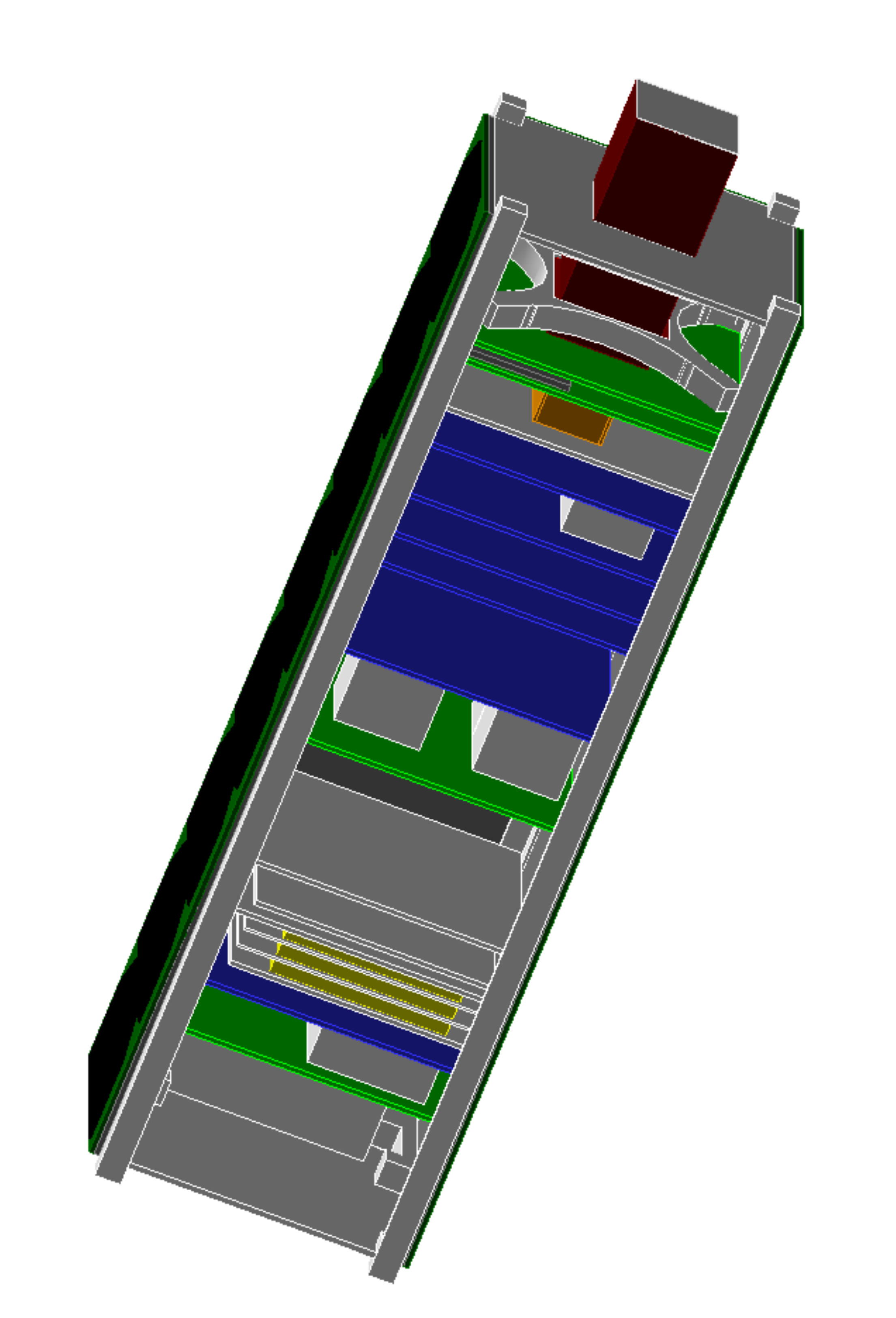}
	\caption{The drawing of Geant4 mass model of Sharjah-Sat-1. The side panel that includes an aluminum plate, solar panels, and corresponding PCBs is not shown to reveal inner structure. }
	\label{fig:SharjahSat1_Mass_Model}       
\end{figure}

Monte Carlo-based simulation toolkit, Geant4, was employed for the background radiation simulations. The Geant4 mass model of Sharjah-Sat-1 used in simulations is shown in Fig.~\ref{fig:SharjahSat1_Mass_Model}. For the geometric mass model, all crucial components are included, such as the CdZnTe crystal, tungsten collimator and tungsten back-shield underneath the crystal, optical light blocker, PCBs, aluminum shielding boxes for the electronic components, batteries, and all structural supports. Geant4 offers a wide range of physics processes and models. Choosing a physics list (physics models for particle interactions with matter in Geant4) that is appropriate for the required application is very important. In our work, we implemented the Shielding Physics List\footnote{\url{https://www.slac.stanford.edu/comp/physics/geant4/slac\_physics\_lists/shielding/physlistdoc.html}} in the simulations. This list consists of all necessary electromagnetic and hadronic physics processes required for the simulations of the background radiation environment in outer space. For each background component, position and energy information of the particle interactions in the crystal, named as events, were recorded for further analysis. 

\subsection{Primary particle generation}

Together with the mass model and the physics lists, we need to determine the spectral energy distributions, positions, directions, and orientations of the primary particles in order to perform simulations in Geant4. We used General Particle Source (GPS) class\footnote{\url{https://geant4-userdoc.web.cern.ch/UsersGuides/ForApplicationDeveloper/html/GettingStarted/generalParticleSource.html}} for the primary particle specifications. For each background component, the spectral models described in Section \ref{sec:background_components} were given as inputs in Geant4. Since the astrophysics sources can be considered as distant point sources, particles emerging from these sources were handled as a parallel beam that envelopes the satellite geometry. On the other side, primary particles created for the background components were emitted inward from a virtual sphere with radius R$_{S}$ with a cosine-law angular distribution. The radius of the sphere was set to 90 cm in the simulations in order to maintain the cosine distribution of the primary particles. The positions and the directions of the primary particles on the sphere were calculated randomly by Geant4. The mass model of the cubesat was located at the center of the sphere and the top of the collimator was considered to be pointing the zenith. The CDGR photons and the GCRs are considered to be coming from outer space, so they were radiated from the upper part of the sphere that covers a solid angle of 8.64 sr. Regarding the altitude between 500-600 km, approximately 3.93 sr is occulted by the Earth's atmosphere. For this reason, the albedo particles and photons were emitted from the bottom part of the sphere which encases a solid angle of 3.93 sr. In the case of the trapped protons and electrons, the particles were radiated the satellite from the solid angle of 4$\pi$ sr.\\

\subsection{Normalization}

In order to calculate the detected count rates, we normalized the simulation results as follows \cite{Sarkar2011}, \cite{Castro2016}. The energy spectrum for each background component was divided into N number of bins equal in logarithmic scale. For each energy bin width, E$_{j}$ ( j is from 1 to N), we created N$_{p}$ primary particles, from a surface area of A, over a solid angle $\Omega$ subtended by the satellite. For each component, N$_{p}$ is considered individually in order to obtain good enough statistics. Since the differential particle flux unit is defined as particles per cm$^{2}$ per second per steradian per keV, particle rate ($\Rho$ - particles$\backslash$s) for each energy bin E$_{i}$ can be calculated by integrating the differential flux as:

\begin{equation*}
	\Rho_{j} = 
	\int dA \int d\Omega \int \frac{f\big(E\big)}{dE} dE_{j}  \qquad  \qquad    (particles \quad s^{-1})
\end{equation*}

where dA =  $R_{S}^{2}$$\int_{0}^{2\pi} d \phi$$\int_{0}^{\theta}sin\theta^\prime d\theta^\prime$. The angle $\theta$ depends on the area of the spherical surface that the particles are radiated from. $\Omega$ is the solid angle of a cone that its apex is the vertex of the incoming particle. 
d$\Omega$ = $\int_{0}^{2\pi}$d$\Phi$ $\int_{0}^{\delta} cos\delta^\prime sin \delta^\prime d \delta^\prime$ and $\delta$ angle ranges from 0 to $\pi$/2. In order to decrease the CPU time and to increase the simulation statistics, one can use a smaller $\delta$ angle less than $\pi$/2.
\\
As a result, the observation time in second, $T$, is obtained by dividing the $N_{p}$ by the calculated particle rate per energy bin.
\begin{equation*}
	T_{j} = 
	\frac{N_{p}}{P_{j}} 
\end{equation*}
The total detected count rate, $C$, is calculated by the summation of the ratio of the number of the deposited particles per energy bin, $M_{j}$, to the observation time $T_{j}$.
\begin{equation*}
	C = \sum_{j=1}^{N}
	\frac{M_{i}}{T_{j}} 
\end{equation*}

\subsection{Sensitivity Calculations}

Sensitivity is one of the most important parameters of a detector system defined as the limiting intensity of the detector for a weak source observation at a particular significance level. It depends on the background radiation and detector properties such as its effective area, energy resolution, observation time, and dead time. The continuum sensitivity can be estimated with the following formula \cite{MATTESON_1978}. 

\begin{equation*}
	F_{min}=SNR\cdot \frac{\sqrt{2.33 \cdot B(E)}}{f_{live} \cdot A_{eff}(E)\cdot \sqrt T \cdot \sqrt{FWHM(E)}}
\end{equation*}  \\

where $ F_{min} $ is the sensitivity of the iXRD at a given energy $E$, $SNR$ is the signal-to-noise ratio, $B$ is the simulated background radiation counts cm$^{-2}$ s$^{-1}$ keV$^{-1}$ at a given energy $E$, $f_{live}$ is the livetime fraction, $A_{eff\ }$ is the effective area of the iXRD at a given energy $E$, $T$ is the total observing time and $FWHM$ is the full width at maximum for the energy resolution of the iXRD at a given energy $E$.

For a realistic sensitivity estimation, one has to consider the iXRD spectral response which depends on the crystal's internal properties, the transportation of the charge carriers created by the source particles, the electric field configuration inside the crystal, and the readout electronic noise. For this reason, we conducted simulations of the charge transportation inside the CdZnTe by using a C++ based simulator THEBES (Transporter of Holes and Electrons By Electric field in Semiconductor Detectors) that was developed by our group. The details of the THEBES simulator can be found in \cite{Altingun2022}. The total background spectrum, B(E), used in the sensitivity calculations was obtained by employing the Geant4 simulations and the THEBES simulator together.


\section{Results}
\label{sec:Results}
In the following, the optimization study for the back-shield design, the background simulation results, count rates for each background component, and sensitivity estimation results are presented.

\subsection{Back-Shield Design}
\label{sec:midshield_design}
The Tungsten collimator enclosing the CdZnTe crystal helps to minimize the effects of the background particles coming from outer space. In addition to that, the simulations indicate that another shield is required, especially to reduce the effect of albedo radiation from the atmosphere. As it is illustrated in Fig.~\ref{fig:ixrdCAD}, the readout electronics design comprises two PCBs, the motherboard and the daughterboard where the daughterboard carries the noise-sensitive analog readout circuitry, and the motherboard carries most of the digital and communication circuitry \cite{Kalemci2022}. A metal plate acting as a passive shielding (called the back-shield) has been placed between those two PCBs. Due to the proximity of the PCBs, we were limited to using a back-shield with a maximum thickness of 2 mm. To achieve optimal performance, we conducted simulations to quantify background rates due to albedo photons by using several materials (W, Pb, Sn, Cd, Cu). The most promising ones out of many configurations are a lead (Pb) shield, a Tungsten (W) shield, and three graded-z shieldings. The results are presented in Table \ref{tab:backshield_counts}.

Making use of the back-shield considerably reduces albedo photon count rates. For the low energy range (20 - 60 keV), all material configurations yield almost similar count rates. However, the 2 mm thick Tungsten provides the lowest counting rate for the total energy range of the iXRD. Tungsten was preferred as the back-shield, because it not only shows good background reduction performance but also was relatively easy to procure compared with the graded-Z configurations.


\begin{table}[ht]
	\begin{center}
		\begin{minipage}{\textwidth}
			\caption{The fraction of count rates for the albedo $\gamma$-rays for different back-shield designs. For each shielding design, the fractions are given as the count rates obtained using the corresponding shielding divided by the count rates with no shield. The thickness of the back-shield is 2 mm. For the graded-Z shielding designs, the thicknesses are 0.5 mm, 1.0 mm, and 0.5 mm respectively.}\label{tab:backshield_counts}%
			\begin{tabular*}{\textwidth}{@{\extracolsep{\fill}}lcccccc@{\extracolsep{\fill}}}
				\toprule%
				\textbf{Energy Range} & \textbf{W} & \textbf{SnWCd}& \textbf{Pb}& \textbf{SnWCu}& \textbf{PbSnCu}\\
				\midrule
				20 - 60 keV & 0.16 & 0.17 &0.16 & 0.18 & 0.21\\
				\\
				20 - 200 keV & 0.14 & 0.16 & 0.16 & 0.17 & 0.19 \\
				\botrule
			\end{tabular*}
		\end{minipage}
	\end{center}
\end{table}

\subsection{Total background spectrum and count rates}
\label{sec:tot_spectrum}

\begin{figure}[ht]
	\includegraphics[width=1.0\textwidth]{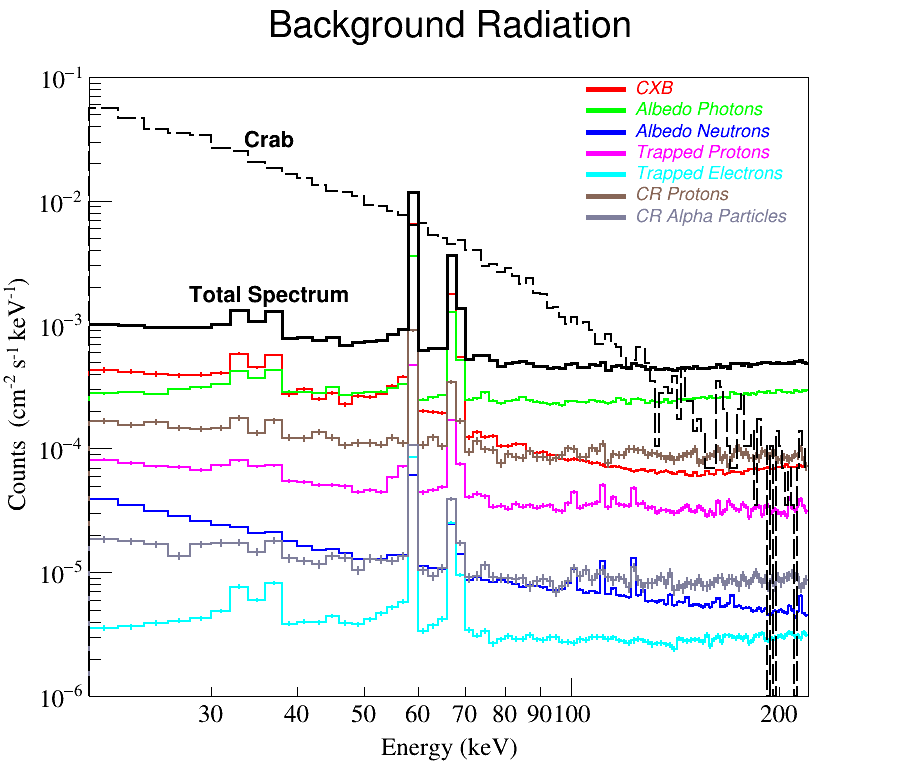}
	\caption{Spectra of the background radiation components obtained with the help of the deposited energies of the background particles in the crystal calculated by Geant4 simulations. The total background is shown and labeled in black color. The crab spectrum is also included in the plot and labeled in black. Tungsten is used as the back-shield material in the simulations.}
	\label{fig:total_bckg}       
\end{figure}

In order to obtain the background in orbit, seven background components were considered (see Section~\ref{sec:background_components}). Fig.~\ref{fig:total_bckg} shows the simulated background spectra obtained using the deposited energies of the background particles in Geant4 simulations. It is worth noting that the detector's spectral response is not included in the background spectra shown. The Crab spectrum is also shown. A power law distribution with parameters $\Gamma$ = 2.08 and N = 8.97 (photons cm$^{-2}$ s$^{-1}$ keV$^{-1}$) was used to model the Crab spectrum with the energy range between 10 keV and 1 MeV. \cite{Kirsch2005}. The CDGR, albedo photons, and the CR protons are the most dominant components. The main contributor to the total background in the energy range of interest (20 - 200 keV) is the albedo photon radiation, which corresponds to roughly $\sim$40$\%$ of the total flux. The energy distribution of the albedo photons is almost flat in the respective energy range. Also, the CDGR and the GC protons significantly contribute to the total background by around $\sim$30$\%$ and $\sim$25$\%$, respectively. Between $\sim$20 and $\sim$60 keV, for which a larger number of source photons are expected due to typical power-law spectra of hard X-ray sources, the contribution of the CDGR photons to the total background is the largest. However, beyond $\sim$70 keV, the CDGR becomes less important than the other most dominant components, the albedo photons and the CR protons. Finally, one can see that the contributions of the albedo neutrons, galactic cosmic alpha particles, and trapped particles are insignificant. In the case of the trapped protons and electrons, the particle fluxes inside the SAA and the geometric polar regions are excluded (see Section~\ref{sec:trped_ch}). Further, the characteristic X-ray emissions from the tungsten (K-line emissions at 57.9 keV, 59.3 keV, and 67.2 keV) are very significant in the background spectra as well. There are also three distinct lines due to the gamma-ray emissions of the excited tungsten nuclei as a result of inelastic neutron and proton scattering in the spectrum, between 100 keV and 130 keV\cite{Lister1967}, \cite{Kruse1971}, but their contribution is insignificant in the overall spectrum. 

The background rates are given in Table~\ref{tab:cnt_rates} for each component. The total count rate (for the particles deposited energies greater than 20 keV) is $\sim$ 7 counts/s. For the energy range of the iXRD payload (20 - 200 keV), the count rate becomes  $\sim$0.85 counts/s, while the rate for the particles with energies between 20 keV and 60 keV is approximately 0.36 counts/s. The count rates for the trapped protons and electrons are also calculated (Table~\ref{tab:cnt_rates_trp}) for the SAA and the polar regions both included and excluded. One can see that the particle count rates are extremely large inside those areas for operating the iXRD. Excluding the high particle-populated regions reduces the trapped particle rates significantly. However, the fact that the iXRD will be turned off during the passage through the SAA and the polar regions will result in a break in scientific observations. The period for one orbit is around 90 minutes ($\sim$15 orbits per day). The time period that the cubesat spends inside the SAA and the polar regions varies between 30 and 40 minutes. This provides a time frame of approximately one hour for the iXRD to operate per orbit. Since the power, telemetry, and control constraints allow around 10-minute operation of iXRD in each orbit, this break can be managed.

\begin{sidewaystable}
	\sidewaystablefn%
	\begin{center}
		\begin{minipage}{\textheight}
			\caption{The background count rates for the corresponding background components in the orbit. The count rates are calculated by using the deposited energies of the background particles in Geant4 simulations.}\label{tab:cnt_rates}
			\begin{tabular*}{\textheight}{@{\extracolsep{\fill}}lcccc@{\extracolsep{\fill}}}
				\toprule%
				\textbf{Energy Range} & \textbf{CDGR} & \textbf{Albedo $\gamma$}& \textbf{GCRs $p^+$}& \textbf{GCRs $\alpha$}  \\
				& \text{(0.01-100 MeV)} & \text{(0.01-100 MeV)}& \text{(0.001-100 GeV)}& \text{(0.001-100 GeV)}\\
				& \text{(cnt/s)} & \text{(cnt/s)}& \text{(cnt/s)}& \text{(cnt/s)}\\
				\midrule
				$\geq$20 keV & 0.39 & 1.58 & 2.70 & 0.30 \\
				20 - 60 keV & 0.17 & 0.12 & 0.04 & 0.005\\
				20 - 200 keV & 0.27 & 0.37 & 0.13 & 0.01 \\
				\botrule
				\textbf{Energy Range} & \textbf{Trapped $p^+$} & \textbf{Trapped $e^-$}& \textbf{Albedo $n^0$}& \textbf{Crab} \\
				& \text{(0.1-400 MeV)} & \text{(0.04-7 MeV)}& \text{(10 keV - 1 GeV)}& \text{(10 keV - 1 MeV)}\\
				& \text{(cnt/s)} & \text{(cnt/s)}& \text{(cnt/s)}& \text{(cnt/s)}\\
				\midrule
				$\geq$20 keV & 2.03 & 0.01 & 0.02& 1.10\\
				20 - 60 keV  &0.02 & 0.002 & 0.005& 0.81\\
				20 - 200 keV &0.05 & 0.005 & 0.01& 1.08 \\
				\botrule

			\end{tabular*}
		\end{minipage}
	\end{center}
\end{sidewaystable}

\begin{table}[ht]
	\begin{center}
		\begin{minipage}{174pt}
			\caption{The background count rates for the trapped particles in the orbit inside and outside of the SAA and the polar regions. The count rates are calculated for the particles that deposit energy greater than 20 keV in the detector volume.}\label{tab:cnt_rates_trp}%
			\begin{tabular}{@{}lll@{}}
				\toprule
				\textbf{Region} & \textbf{Trapped $p^+$} & \textbf{Trapped $e^-$}\\
				&  \text{(0.1-400 MeV)} & \text{(0.04-7 MeV)}\\
				& \text{(cnt/s)}& \text{(cnt/s)} \\
				\midrule
				Including \\ SAA and Polar Regions & 9047.7 & 1240.0 \\
				\\
				Excluding\\ SAA and Polar Regions & 2.03 & 0.02 \\

				\\
				\botrule
			\end{tabular}
		\end{minipage}
	\end{center}
\end{table}

\subsection{Sensitivity}
\label{sec:sensitivity}

In order to estimate the sensitivity curve, we performed THEBES simulations using the Geant4 background simulation outputs as input. The positions and the energy deposition of the electron-hole pairs from Geant4 were imported into the THEBES simulator to calculate the induced signals per background event. 
To validate the dedicated Geant4 and THEBES simulations, a comparative study between an experiment and the corresponding simulation set was conducted. We carried out a set of experiments with $^{57}$Co (122.1 keV, 136.5 keV) source for on-ground calibration of the energy response of the iXRD. The iXRD system was housed in a metal box. The radioactive source was then placed on the top of the box, facing the top of the collimator. The details of the calibration setup are reported in \cite{Kalemci2022}. More than 5 $\times 10^{5}$ events were registered in all channels.

\begin{figure}[ht]
	\includegraphics[width=0.8\textwidth]{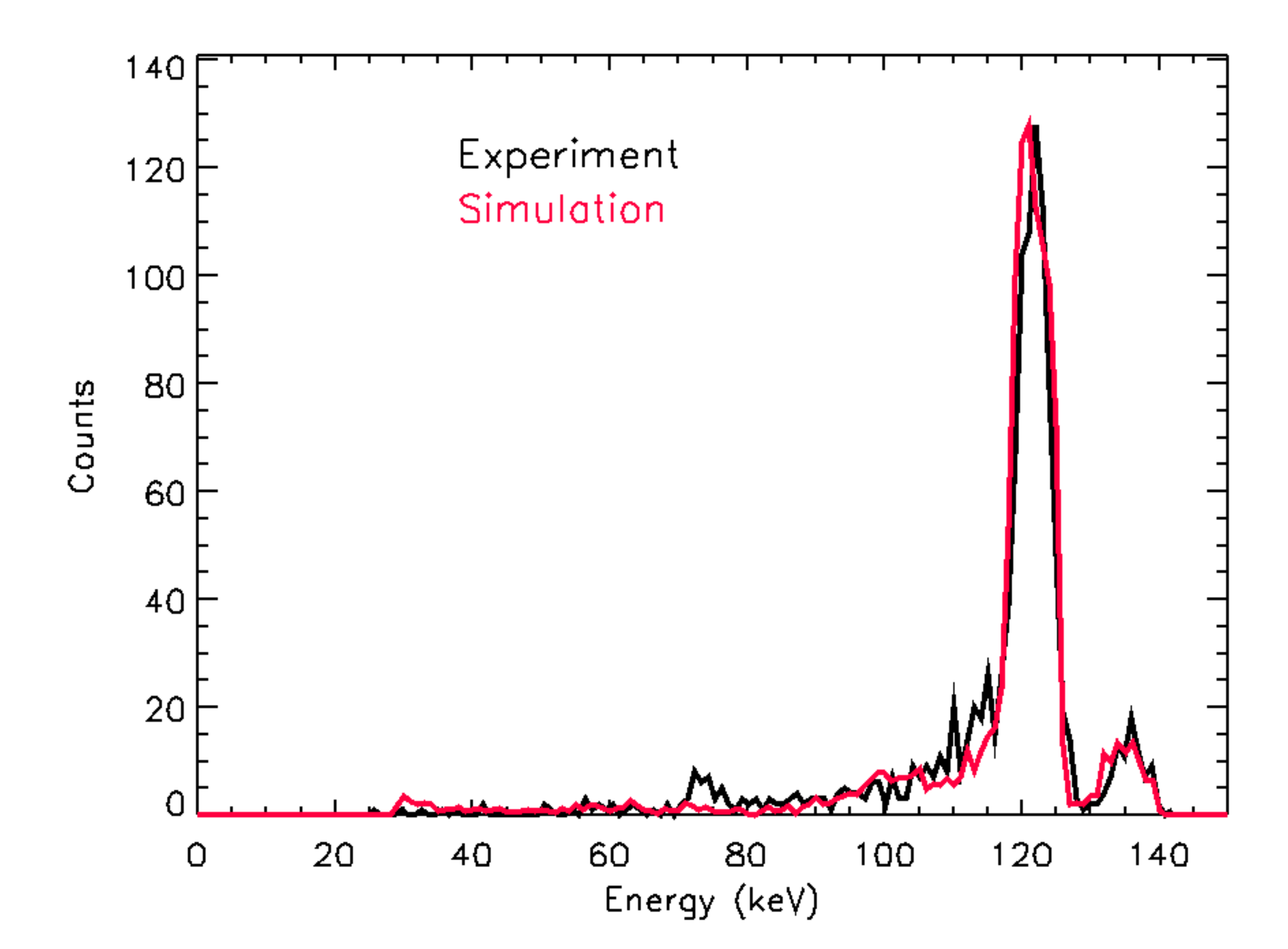}
	\caption{The comparison of the energy spectra for the simulations and experiments for a single channel. For comparative purposes, the simulation result is normalized by utilizing the 122 keV peak count from the measurements.  }
	\label{fig:sim_vs_exp}       
\end{figure}

In the meantime, we modeled the experimental setup in Geant4 as well. The Shielding Physics List was chosen and approximately 4.5 $\times 10^{5}$ events were obtained using a $^{57}$Co source. The positions and the deposited energies of the electron-hole pairs created by the incident particles were registered. The induced signals were calculated with the THEBES simulator for four different channel groups as a final step.   Fig.~\ref{fig:sim_vs_exp} illustrates an example from the comparison results for a single channel. The electronic noise level for the single pixel was assigned as 4\% and for the planar cathode, the noise level was 6\%. The simulations are in good agreement with the measurements.\\ 
The THEBES simulation parameters that provided the best simulated $^{57}$Co spectra compared to the measured ones are given in Table \ref{tab:sens_sim_parameters}. 

\begin{table}[ht]
	\begin{center}
		\begin{minipage}{174pt}
			\caption{The simulation parameters to obtain the iXRD background spectrum.}\label{tab:sens_sim_parameters}%
			\begin{tabular}{@{}ll@{}}
				\toprule
					\textbf{THEBES Parameters} & \textbf{Values} \\
				\midrule
				Electron mobility (mm$^{2}$/V s)& 1.0 $\times 10^{5}$ \\
				Hole mobility (mm$^{2}$/V s) & 1.05 $\times 10^{4}$\\
				Electron trapping time (s) & 3.0 $\times 10^{-6}$\\
				Hole trapping time (s) & 1.0 $\times 10^{-6}$\\
				Electronic noise level for anode channels & 4-7\%\\
				Electronic noise level for planar cathode & 6\%\\
				Minimum detectable energy for anodes & 20 keV\\
				Minimum detectable energy for cathode & 40 keV\\
				\botrule
			\end{tabular}
		\end{minipage}
	\end{center}
\end{table}

\begin{figure}[ht]
 	\includegraphics[width=1.0\textwidth]{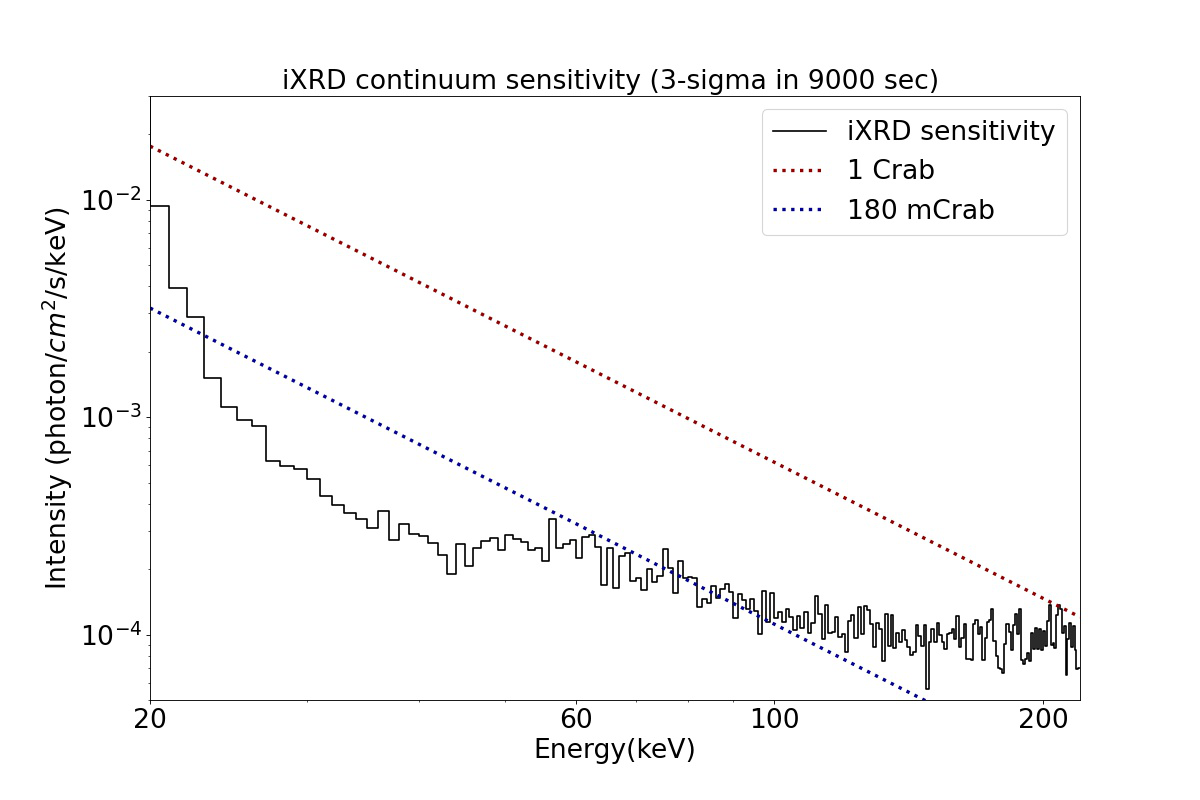}
	\caption{Simulated sensitivity curve of iXRD with an observation time of 9000 seconds, a live-time fraction of 90\% and a SNR of 3. }
	\label{fig:sensitivity}       
\end{figure}

In Fig.~\ref{fig:sensitivity}, the calculated sensitivity curve of the iXRD is presented with the Crab spectrum. We considered a background simulation of 9000 seconds of integration time (indicating one day) due to the 600 seconds operation time of the iXRD in each orbit. The induced background radiation signals for all channel groups were then calculated in the THEBES simulator. According to the experimental data, the noise levels were set to be 4\%, 5\%, 6\%, and 7\% for the single, small, medium, and large channels, respectively. The effective area of the iXRD was simulated using a circular source radiating a mono-energetic beam of photons ranging from 10 keV to 300 keV. Finally, a $SNR$ of 3, a live-time fraction of 0.9, and simulated energy resolution values for the energy range of 20 keV to 240 keV were considered in the sensitivity calculations.

The simulations showed that the continuum sensitivity of the iXRD is about 180 mCrab ($\sim$0.07 photons cm$^{-2}$  s$^{-1}$)  between 20 to 100 keV energy band. This sensitivity is enough to monitor very bright X-ray sources in the Galaxy (such as Cyg X-1 and GRS 1915$+$105).

\section{Discussion and Conclusion }
\label{sec:discussion}
In this paper, we presented a simulation study on the estimation of the effects of the background radiation on the performance of the iXRD, a CdZnTe-based scientific payload of Sharjah-Sat-1 cubesat. The albedo photon radiation is the most dominant and flat in the iXRD energy range, while the CDGR radiation contribution is the highest in the energy range of most of the astrophysical sources, from $\sim$20 keV to $\sim$60 keV. Another dominant component is the galactic cosmic protons, while the radiation effects of the trapped particles, the albedo neutrons, and the cosmic alpha particles are negligible. In addition, we did not consider the secondary protons, electrons, and positrons created by the interaction of the cosmic rays with the atmosphere since their contributions are insignificant.

The radiation environment in LEO has a dynamical nature and the intensity of the components can vary considerably. The albedo photon background contribution depends on the orientation of the cubesat. For the background simulations, the cubesat was oriented to be pointing the zenith, in which the albedo radiation would be quite effective. Orienting the cubesat at an angle of 45$^{\circ}$ relative to the zenith decreases the total count rate for the iXRD energy range because less number of photons are registered due to the tilt of the detector plane with respect to the zenith, although the fluorescence emission increases due to the exposure of the collimator to more albedo photons from the side of the crystal. However, a zenith angle of 45$^{\circ}$ reduces the albedo photon count rate by only around 5$\%$ in the energy band of 20-200 keV. This shows that albedo photons are the most dominant and almost immutable background component. The effect of the cosmic rays depends on the geomagnetic cutoff rigidity. Due to the low cutoff rigidities in the high geomagnetic latitudes, the effect of the charged particles becomes more significant \cite{Liao2020}. Since the operation of the iXRD will be halted during the passage through regions with high latitudes, the cosmic ray count rates will also decrease. Therefore, the worst-case scenario is considered for the effects of the background radiation in this work. Also, the total count rate is around 7 counts/s, which is critical to determine the live-time fraction of iXRD\cite{Kalemci2022}.

This study does not consider the delayed background radiation due to the activation of radioactive isotopes within the satellite material. However, when considering large count rates for the trapped particles (see Table \ref{tab:cnt_rates_trp}) during the passages through the SAA and the polar regions for the anticipated orbit and high-density materials used in the iXRD system, the spectral performance of the iXRD can suffer from the emissions from short-lived induced isotopes \cite{ODAKA201892}. The trapped proton simulations indicate proton-induced radioisotopes with half-lives in the order of minutes such as  $^{109m, 107m, 105m}$Ag, $^{104, 111m}$Cd  formed in the CdZnTe crystal and $^{176}$W, $^{177m, 179m, 183m}$Ta, $^{170m, 172m, 173}$Lu in the collimator and the back-shield. Since in each orbit the operation period of the iXRD is expected to be around 10 minutes and the time period that the iXRD will be outside of the SAA and the polar regions is approximately 1 hour per orbit (see Section \ref{sec:tot_spectrum}), the effects of the delayed emission can be eliminated by activating the iXRD within a reasonable time after leaving those areas. This will be part of the future planning of observations.

The expected detection sensitivity was obtained with the help of the background simulations and THEBES charge transportation simulations. It is around 180 mCrab between 20-100 keV energy band in one day. The sensitivity result will assist us in creating a strategic plan for observations in order to attain the scientific objectives of the iXRD.

Finally, the iXRD was designed and developed from the ground up by our group and the simulation studies have provided us a considerable amount of information about a detector system that can work in the space environment. This great deal of technological know-how will have significant contributions for prospective future projects with extensive science goals.  

\bmhead{Acknowledgments}

The development of iXRD has been supported by the University of Sharjah, Sabanc\i\ University and T{\"u}bitak Project 116F151.

\bmhead{Author Contributions}

 AMA and EK prepared, analyzed, and interpreted simulation and experimental data. EÖ prepared Fig. 11 and interpreted data for sensitivity calculations. The first draft of the manuscript was written by AMA and all authors commented on previous versions of the manuscript. All authors read and approved the final manuscript.
 
\bmhead{Funding}
 This work is supported by the University of Sharjah, Sabanc\i\ University and T{\"u}bitak Project 116F151.
 
\bmhead{Data Availability} The measured and analyzed data of this work are available from the corresponding author upon reasonable request.
 
\bmhead{Code Availability} The code used for this work is custom-made and available from the corresponding author upon reasonable request.

\section*{Declarations}
 \bmhead{Conflicts of interest} The authors have no relevant financial or non-financial interests to disclose.

\bibliography{sn-bibliography}


\end{document}